\documentclass[preprint,showpacs,preprintnumbers,amsmath,amssymb]{revtex4}
\usepackage{graphicx}
\usepackage{dcolumn}
\usepackage{bm}

\def\vec#1{\mbox{\boldmath $#1$}}
\begin{document}

\title{
Velocity Distribution of Inelastic Granular Gas in
Homogeneous Cooling State
}

\author{Hiizu Nakanishi}

\affiliation{Department of Physics, Kyushu University 33, 
Fukuoka, 812-8581, Japan
}
\date{ }

\begin{abstract}
The velocity distribution of inelastic granular gas is examined
numerically on two dimensional hard disk system in nearly elastic regime
using molecular dynamical simulations.  The system is prepared initially
in the equilibrium state with the Maxwell-Boltzmann distribution, then
after a several inelastic collisions per particle, the system falls in
the state that the Boltzmann equation predicts with the stationary form
of velocity distribution.  It turns out, however, that due to the
velocity correlation the form of the distribution function does not stay
time-independent, but is gradually returning to the Maxwellian
immediately after the initial transient till the clustering instability
sets in.  It shows that, even in the homogeneous cooling state, the
velocity correlation in the inelastic system invalidates the assumption
of molecular chaos and the prediction by the Boltzmann equation fails.
\end{abstract}
\pacs{45.50.-j, 51.10.+y, 45.70.-n}

\maketitle


Free cooling of granular gas under no gravity has been attracting much
interest as a subject of statistical mechanics since people realized
that the inelastic collisions between particles makes the system behave
very different from the elastic system, that is the subject of the
conventional statistical mechanics.

Besides the cooling, or loosing its kinetic energy due to the
inelasticity, it has been well recognized by now the system shows series
of instabilities however small the inelasticity may be as long as the
system is large enough.  If the system is
prepared in a highly agitated state, initially it cools down uniformly
as
\begin{equation}
 T = {T_0\over (1+t/t_0)^2},
\label{T-decay}
\end{equation}
which is called Haff state \cite{Haff83}.  After a while, the vortex
structure develop in the velocity field(the shearing
instability)\cite{NE00}; then the uniformity of the particle density is
broken(the clustering instability)\cite{GZ93}.

After the clustering instability, the clusters of particles collide with
each other, merge, and split in a complex way \cite{LH99}; the system
eventually develops high density region, where the inelastic collapse
\cite{MY92,MY94} is likely to happen if one consider ideal hard sphere
system with a constant restitution coefficient.

Regarding the velocity distribution, the Maxwell-Boltzmann distribution
is an equilibrium velocity distribution for the elastic system, and the
relaxation to the distribution is known to be very fast, i.e. within a
several collisions per particles when it is spatially uniform.  In the
case of the inelastic system, obviously any velocity distribution cannot
be stationary because the system loses kinetic energy at every
collision, but it is plausible that the form of distribution stays
stationary after a short transient if the velocity is scaled by the
average speed $v_0(t)$:
\begin{equation}
 f(\vec v, t) = 
{1\over v_0(t)^d}\hat f\left({\vec v\over v_0(t)}\right).
\label{vel-dist}
\end{equation}
In fact, kinetic theories based on the Boltzmann equation predicts that,
after a several collisions per particle, the velocity distribution for
the inelastic system falls into a stationary form that is different from
Gaussian \cite{NE98,BP00,HOB00}. 

In this report, I present results of large scale two dimensional MD
simulations and shows that the form of the velocity distribution does
not stay stationary in the inelastic gas, but after a short initial
transient the distribution gradually getting back to the Gaussian till
the clustering instability sets in.  This gradual change starts at very
early stage where the inhomogeneity in the system is hardly visible.

The system we examine is the two dimensional system of hard disks that
undergo inelastic collisions with a constant normal restitution $r$.
The rotational motion is ignored. Then, the collision rule is given by
\begin{eqnarray*}
\vec v'_i & = &
    \vec v_i - {1+r\over 2}\Bigl(\vec n\cdot (\vec v_i-\vec v_j)\Bigr) \vec n
\\
\vec v'_j & = &
    \vec v_j + {1+r\over 2}\Bigl(\vec n\cdot (\vec v_i-\vec v_j)\Bigr) \vec n,
\end{eqnarray*}
where $\vec v_i$ and $\vec v'_i$ are the velocity
of the $i$'th disk before and after the collision with the $j$'th
particle, respectively, and $\vec n$ is the unit vector parallel with
the relative position of the colliding particles at the time of contact.

The average speed $v_0(t)$ for $d$-dimensional system, defined by
\begin{equation}
 {d\over 2}v_0(t)^2 = \int d\vec v f(\vec v, t) \vec v^2,
\end{equation}
decreases as the system looses energy; with this speed we scale the
velocity distribution through eq.(\ref{vel-dist}).
In order to see the time-dependence of
the scaled velocity distribution $\hat f$,
it is convenient to expand it using the
Sonine polynomial as
\begin{equation}
 \hat f(\vec c,t) = {1\over\sqrt{\pi}^d} e^{-c^2}\sum_{\ell = 0}^\infty
a_\ell(t)S_\ell(c^2)
\end{equation}
when the distribution is not very different from Gaussian; the $\ell$'th
order Sonine polynomial is the $\ell$'th order polynomial orthogonalized
with the $d$-dimensional Gaussian weight function:
\[
 S_0(x)=1,\, S_1(x)=-x+{1\over 2}d, \,
\]\[
 S_2(x) = {1\over 2}x^2-{1\over 2}(d+2)x+{1\over 8}d(d+2), \mbox{ etc.}
\]
Due to the normalization and scaling of $\hat f$, we have $a_0=1$ and
$a_1=0$, thus any deviation from Gaussian distribution is seen in the
non-zero values of $a_\ell$ for $\ell\ge 2$.


Simulations were performed by the event-driven method using the fast
algorithm developed by Isobe \cite{I99}.  Most of the simulations were done with
particle number $N=$250,000, number density $n=0.25$ (area fraction
$\phi\equiv\pi n/4=0.196$).  I employed the periodic boundary condition and the
initial state is the equilibrium state that is prepared by running the
system for long enough with the restitution constant $r=1$.  We focus on
the nearly elastic regime where the system stays uniform for a
substantial length of time and the distribution does not deviate very
much from the Gaussian when the system is in HCS.

In the following, the time is measured by the collision time $\tau$,
which is defined as the number of collisions each particle experiences,
i.e.  $\tau\equiv 2N_{\rm coll}/N$ with $N_{\rm coll}$ being the total
number of collisions (the factor 2 comes from the binary collision).


Figure 1 shows the energy decay as a function of time $\tau$ for
$r=0.9$ and $n=0.25$ in the semi-logarithmic scale.  
In terms of $\tau$, the decay in the Haff state given by (\ref{T-decay}) is
expressed as
\begin{equation}
E(\tau) = E(0) \exp[-\gamma\tau],
\label{E-decay}
\end{equation}
where $\gamma$ is a decay rate.  The thin solid line in Fig.1 shows the
exponential function with $\gamma=0.093$(the line is shifted vertically
to avoid complete overlapping).
The initial $\tau$-dependence fits to the exponential decay very well
with the decay rate very close to the one obtained in the case of random
collision: $\gamma_0\equiv (1-r^2)/d=0.095$ for $r=0.9$.  It eventually
deviates from the exponential around $\tau\sim 70$, when the clustering
instability sets in.

The speed distribution for this system is plotted in Fig.2 for $\tau=40$
and 80.  For both cases, the distribution is very close to Gaussian and
the deviation from it is hardly seen.

The deviation, however, is clearly seen in $a_2(\tau)$ plotted in Fig.3,
where the initial deviation from the Gaussian is shown for various
values of restitution constant $r$.
From this figure, it might seem that the scaled distribution becomes
stationary after a several collisions per particle as is expected from
the kinetic theories \cite{NE98,BP00,HOB00}.  These ``stationary''
values of $a_2$ agree very well with the results of kinetic theory for
the nearly elastic region $r\ge 0.95$ (Fig.4).

This form of distribution, however, is not really stationary as it may
look in the initial stage data of Fig.3.  The $\tau$ dependence of
Sonine coefficients over longer time scale is shown in Fig.5 for
$r=0.9$, 0.95, and 0.98.  In this time scale, the plateau is hardly seen
and absolute value of all the coefficient show gradual decrease towards
zero till the time when the clustering instability sets in; after that
time the distribution deviates from Gaussian drastically.

The Sonine coefficient $a_2(\tau)$ for various $r$ is plotted in Fig.6
where the time is scaled by the clustering time $\tau^*$ when the
clustering instability sets in and $a_2$ is scaled by its
maximum absolute value $|a_2^*|$ for each $r$. 
The closer the value of $r$ is to 1, the smaller
the slope in the $\tau$ dependence becomes, but for all cases, the gradual
return to Gaussian starts almost immediately after the initial transient
period finishes.  It starts actually far before any instability becomes
evident.

This behavior obviously contradicts to the results of the kinetic
theories based on the Boltzmann equation \cite{NE98,BP00,HOB00}; the
theories predict the distribution shows the stationary form after the
short initial transient.  The stationary form should last till the
Boltzmann equation becomes invalid due to correlations developed in the
system.  The fact that the distribution function starts to deviate from
the stationary form quite early stage suggests that the correlation due
to inelasticity becomes important much earlier than it is generally
expected \cite{com1}.


The correlation that is responsible to the behavior of the velocity
distribution is the velocity correlation.  This can be seen by examining
the behavior of the system where the particle velocity is artificially
randomized by the operation that the velocity of each particle is
shuffled by exchanging them between pairs chosen randomly.  By doing
this, we destroy the spatial correlation of velocity while preserving
the velocity distribution.  The dashed line in Fig.1 shows the energy
decay with the velocity shuffling and indicates that the clustering is
prevented by the velocity shuffle.  From Fig.7, we can see clearly that
this system shows the stationary form of the velocity distribution whose
Sonine coefficients are closer to those predicted by the theories.

In Fig.7, the density dependence of the Sonine coefficient behavior is
also examined for $n=$0.25, 0.111, and 0.0625.  The general tendency
that $|a_2|$ decreases toward zero after the initial deviation is the
same and they are all clearly different from the case with the velocity
shuffling although a certain density dependence exists in the present
density region\cite{com2}.


The kinetic theories have succeeded in explaining many aspects of
granular gas, but the Boltzmann-Enskog equation, on which most of the
theories based, ignores the particle correlations except for the pair
correlation factor of the position.  This approximation is quite good in
the equilibrium system thanks to the absence of the velocity
correlation.  In the inelastic systems, however, the system develops the
velocity correlation, and
the assumption of molecular chaos fails even at very early stage,
where the system is still in HCS.
This invalidates the prediction based on the Boltzmann equation that the
functional form of the velocity distribution of the inelastic system in
HCS becomes stationary.

In summary, using large scale MD simulations, I have demonstrated that
in the inelastic system the velocity distribution does not stay in a
stationary form, contrary to the expectation by the kinetic theories
based on the Boltzmann equation.  This is due to the velocity
correlation developed through the inelastic collisions, and this effect
manifests itself in the velocity distribution from very early stage
where any instabilities caused by the inelasticity are still hardly
visible.


\newpage

\begin{figure}
\begin{center}
\includegraphics[angle=0,width=8.cm]{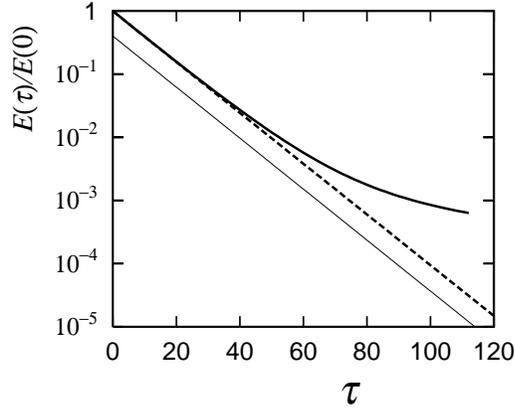}
\end{center} %
\caption{\label{fig-1} Energy decay as a function of $\tau$ for $r=0.9$
and $n=1/4$.  The solid line denotes the result for the system with the
ordinary inelastic dynamics and the dashed line for the system with the
velocity shuffle(see the text).  The exponential decay (5) with
$\gamma=0.093$ is also plotted by the thin solid line with an extra
factor to avoid complete overlapping with the dashed line.  }
\end{figure}
\newpage

\begin{figure}
\begin{center}
\includegraphics[angle=0,width=8.cm]{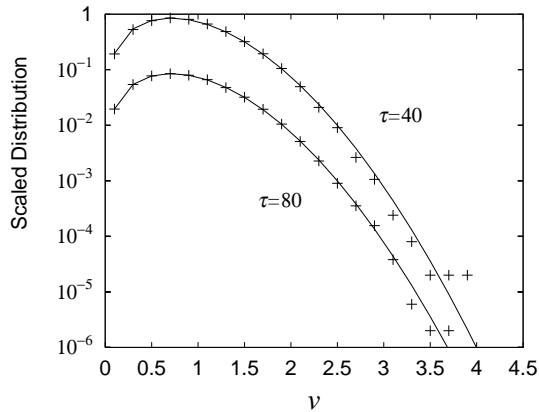}
\end{center} %
\caption{\label{fig-2} 
The scaled speed distributions for $\tau=$40 and 80 with $r=0.9$ and $n=1/4$.
The plot for $\tau=80$ is shifted by the factor $10^{-1}$.
The Maxwell-Boltzmann distributions are indicated by the solid lines for
 comparison.
}
\end{figure}

\newpage

\begin{figure}
\begin{center}
\includegraphics[angle=0,width=8.cm]{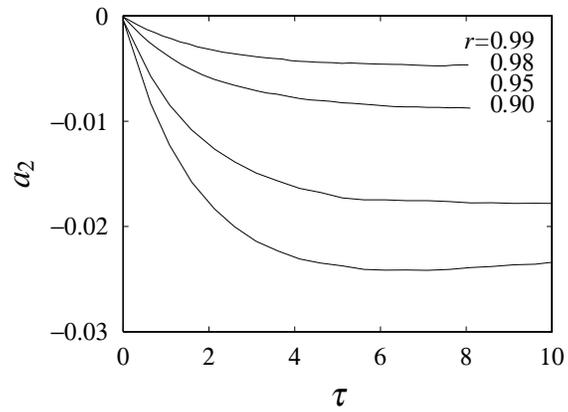}
\end{center} %
\caption{\label{fig-3} The initial time dependence of $a_2$ for
$r=$0.99(top), 0.98, 0.95, and 0.90(bottom) with $n=1/4$.
}
\end{figure}

\newpage

\begin{figure}
\begin{center}
\includegraphics[angle=0,width=8.cm]{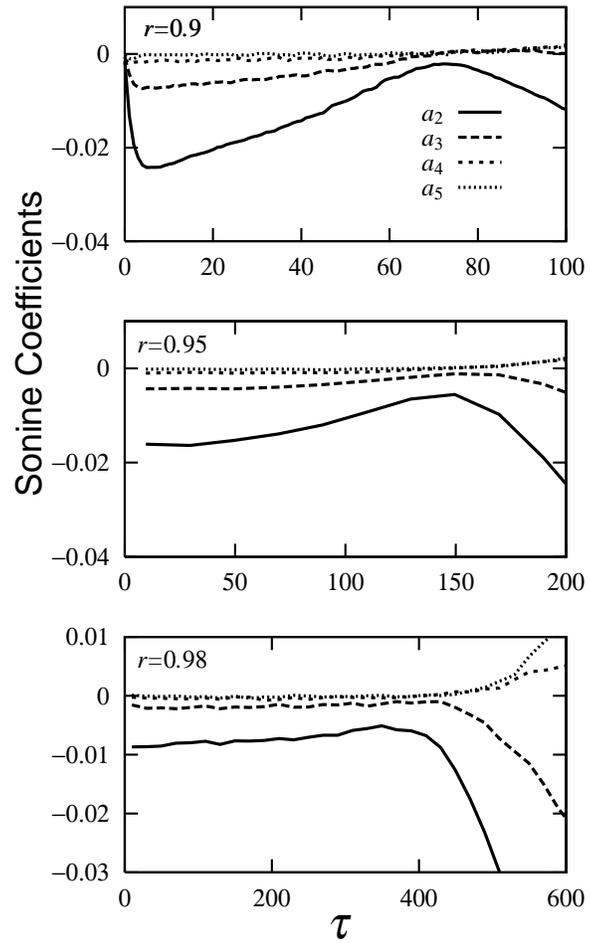}
\end{center} %
\caption{\label{fig-4} 
The Sonine coefficients $a_\ell$ ($2\ge\ell\ge 5$) for $r=$ 0.9(top),
 0.95(middle), and 0.98(bottom) as functions of $\tau$ for $n=1/4$.
}
\end{figure}
\newpage

\begin{figure}
\begin{center}
\includegraphics[angle=0,width=8.cm]{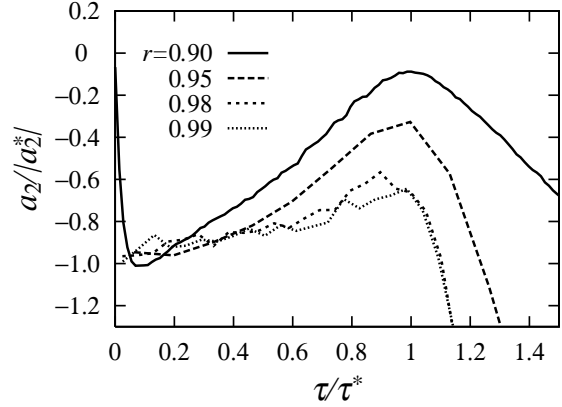}
\end{center} %
\caption{\label{fig-5} 
The scaled Sonine coefficient $a_2/|a_2^*|$ v.s. the scaled time
 $\tau/\tau^*$ for $r=$0.90, 0.95, 0.98, and 0.99 with $n=1/4$.
}
\end{figure}
\newpage

\begin{figure}
\begin{center}
\includegraphics[angle=0,width=8.cm]{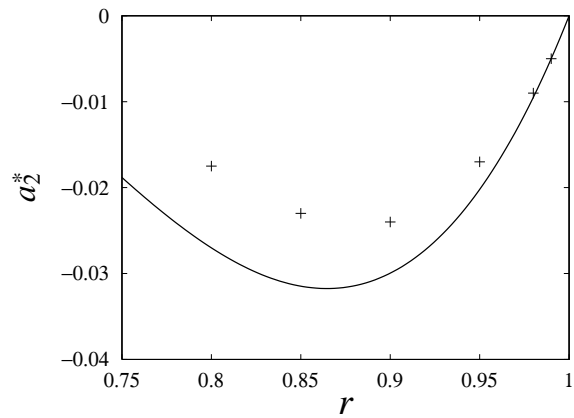}
\end{center} %
\caption{\label{fig-6} 
The minimum values for $a_2$ v.s. the restitution constant $r$.
The solid line indicates the results by the Boltzmann equation\cite{HOB00}.
}
\end{figure}
\newpage

\begin{figure}
\begin{center}
\includegraphics[angle=0,width=8.cm]{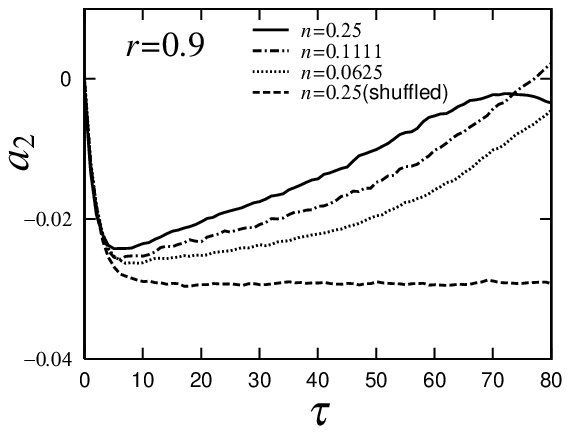}
\end{center} %
\caption{\label{fig-7} The Sonine coefficient $a_2$ for $r=0.9$ as a
function of $\tau$ for the ordinary dynamics (solid line) and the
velocity shuffled dynamics (dashed line) for $n=1/4$.  $a_2$ for
$n=0.1111$ (the dash-dotted line) and $n=0.0625$ (the dotted line) with
the ordinary dynamics are also plotted.  }
\end{figure}



\begin{thebibliography}{99}
\bibitem{com2}
The behavior after the clustering($\tau\ge 70$) shows strong sample
	dependence for the present system size ($N=250,000$) while
	the behavior in the homogeneous state is quite stable.
\end{thebibliography}
\end{document}